\documentclass[english]{article}
\usepackage[T2A,T1]{fontenc}
\usepackage[latin9]{inputenc}
\usepackage{geometry}
\usepackage{amsmath}
\usepackage{graphicx}

\makeatletter
\newcommand{\lyxaddress}[1]{
\par {\raggedright #1
\vspace{1.4em}
\noindent\par}
}

\@ifundefined{date}{}{\date{}}
\usepackage{babel}

\usepackage{babel}

\usepackage{babel}

\usepackage{babel}

\usepackage{babel}

\@ifundefined{showcaptionsetup}{}{%
 \PassOptionsToPackage{caption=false}{subfig}}
\usepackage{subfig}
\makeatother

\usepackage{babel}
\begin{document}

\title{A learning rule balancing energy consumption and information maximization
in a feed-forward neuronal network}

\author{Dmytro Grytskyy$^{1}$, Renaud Blaise Jolivet$^{1,2}$}
\maketitle

\lyxaddress{$^{1}$\textit{Département de Physique Nucléaire et Corpusculaire
(DPNC), University of Geneva}, Geneva, Switzerland}

\lyxaddress{$^{2}$\textit{CERN}, Geneva, Switzerland}

\lyxaddress{e-mail: dmytro.grytskyy@etu.unige.ch or renaud.jolivet@unige.ch}
\begin{abstract}
Information measures are often used to assess the efficacy of neural
networks, and learning rules can be derived through optimization procedures
on such measures. In biological neural networks, computation is restricted
by the amount of available resources. Considering energy restrictions,
it is thus reasonable to balance information processing efficacy with
energy consumption. Here, we studied networks of non-linear Hawkes
neurons and assessed the information flow through these networks using
mutual information. We then applied gradient descent for a combination
of mutual information and energetic costs to obtain a learning rule.
Through this procedure, we obtained a rule containing a sliding threshold,
similar to the Bienenstock-Cooper-Munro rule. The rule contains terms
local in time and in space plus one global variable common to the
whole network. The rule thus belongs to so-called three-factor rules
and the global variable could be related to a number of biological
processes. In neural networks using this learning rule, frequent inputs
get mapped onto low energy orbits of the network while rare inputs
aren't learned. 
\end{abstract}

\subsection*{Keywords}

Mutual information; energy consumption; plastic networks; non-linear
Poisson neurons; three-factor rules; noise

\section{Introduction}

It is typically assumed that neural networks perform some form of
information processing. It is thus a natural idea to apply information
theory to assess the efficacy of those networks. Various mutually
related information measures have been used in this context. For instance,
Kramers-Rao relation and Fisher information are used in reference
\cite{BrunelNadal} to obtain an upper limit for information inference,
and to relate it to the mutual information between a network's inputs
and outputs in the limit of a large number of neurons. The authors
then search for the form of a neuron's non-linear transmission function
optimizing this mutual information measure. In reference \cite{Nadal94},
the authors similarly search for the optimal form of neuronal non-linearity
using the Kullback-Leibler divergence. It is also natural to think
of synaptic plasticity induced by neuronal activity as geared towards
increasing the performance of information inference. In particular,
it is possible to derive learning rules converting neuronal activity
in synaptic changes as an optimization procedure of some information
metric. In reference \cite{Gerstner04} for instance, Fisher information
is applied to link the refractory period after a spike has been elicited
to the shape of the Spike-Timing Dependent Plasticity (STDP) learning
window. Nessler and colleagues \cite{Maass13} have also obtained
learning rules approximating expectation maximization of the likelihood
measure between inputs generated by hidden causes and inputs reconstructed
from neurons' outputs. Reference \cite{Maass14} further generalizes
this approach for learning of sequences of inputs in time. In both
works, the authors additionally exploit fast global inhibition to
achieve a winner takes all regime. A likelihood measure expressed
with Kullback-Leibler divergence is also applied in \cite{Brea} to
derive learning rules in an expectation maximization manner for a
recurrent network learning a set of sequences. In \cite{Linsker88},
Linsker formulated an infomax principle, proposing maximization of
information about the presented input in the network's output measured
by mutual information, as a goal for network parameters fitting. This
concept is applied in \cite{Linsker92} for a feed-forward architecture
with linear neurons and noisy input, generalized in \cite{BellSejnovski95}
for non-linear neurons without noise, and applied in \cite{Linsker97}
for recurrent networks of non-linear neurons solving the task of independent
component analysis. In \cite{Stemmler99}, mutual information is used
to formulate a general form for plasticity rules and applied to the
Hodgkin-Huxley model. Finally, in \cite{Chechik}, learning rules
optimizing mutual information between neurons' inputs and outputs
are derived in the approximation of rare appearance of inputs in a
noisy background.

Theoretically, networks can learn an arbitrary complex structure of
external inputs. In biological neural networks however, operational
and computational capacities are not arbitrarily large, but are restricted
by the amount of available resources. Energy is one of these resources
\cite{Jolivet12,Conrad2018}. In the mammalian brain, energy is mostly
spent at synapses, to power post-synaptic potentials, and to a lesser
extent to power action potential generation \cite{Yu_rev,Jolivet2009,Jolivet12,JolivetJCBFM}.
Note that such restrictions are also relevant for artificial neural
networks, particularly for neuromorphic hardware implementations,
for example through restricted electric supply, limited capacity of
communication channels, or limited capacity for thermal cooling. It
is therefore reasonable to investigate optimal information processing
in neural networks in relation to concomitant energetic costs, or
under restrictions on available energetic resources, briefly formulated
as ``bits per joule'' \cite{Levy_long}, or more physiologically,
``bits per ATP'' (Adenosine Triphosphate, the main energy currency
of mammalian cells). Reference \cite{Levi_short} for instance, obtained
results on the optimal intensity of information transmission over
axons. Tsubo and colleagues \cite{Tsubo} obtained a biologically
realistic distribution of interspike intervals considering a network
optimizing the firing rates distribution under competing reliability
and energetic constraints. In reference \cite{Sengupta_simEvsI},
Sengupta and colleagues demonstrated that an optimal mutual information
``bits per joule'' relation is achieved by a moderate relation between
excitation and inhibition. We have also recently demonstrated \emph{in
silico} and in \emph{in vitro} experiments that synapses in the cortex
and in the visual pathway are tuned to maximize ``bits per ATP''
\cite{Jolivet12,Conrad2018,Harris2015,Harris2019}. Similar results
have been obtained by Kostal and colleagues \cite{Kostal2015,Kostal2016}.
In reference \cite{Sengupta_freeE}, Sengupta and colleagues connected
the optimization of mutual information and energetic costs to the
network's free energy via thermodynamic-like considerations, and related
them respectively to accuracy and simplicity. Reference \cite{Yu_NvsE}
obtains the optimal number of neurons coding for a noisy signal depending
of noise intensity and energetic costs. Simulations in \cite{Tkacik}
investigating this question obtain different regimes for different
parameter values. Recently, a sequence of essential works was published,
in which learning rules optimizing bits per joule are derived. A network
generating a spike only if it conveys enough information on the coded
signal to cover its energetic price is investigated in \cite{Deneve_spikes},
deriving learning rules for a spiking recurrent network. This approach
is further applied to rate coding in \cite{Deneve_rates}, and to
oscillating activity reducing energetic costs in networks of noisy
neurons in \cite{Gutkin_osc}.

Here, we first derive learning rules optimizing information inference,
before balancing information processing efficacy \textendash{} measured
with mutual information \textendash{} with concomitant energetic costs.
We then consider a feed-forward network with the derived learning
rule, and obtain a relation between the probability of specific input
patterns and the respective energetic cost of learned evoked responses.

\section{Methods}

We consider a network of non-linear Hawkes neurons, with non-linearity
$f(u)$ mapping a neuron's membrane potential $u$ to its firing probability
$p(y)$, with the neuron's output $y$ a binary variable with $y=1$
for spikes and $y=0$ for no spikes. The membrane potential $u$ evolves
in time following a leaky integrating dynamics: 
\begin{equation}
d_{t}u=-(u-u_{0})/\tau_{m}+I,
\end{equation}
so that: 
\begin{equation}
u(t)=[h*I](t)=\int_{-\infty}^{^{t}}h(t-t_{1})I(t_{1})dt_{1}
\end{equation}
with $*$ standing for convolution with the kernel function $h(\tau)$.
Here, $u_{0}=0$ is used for the sake of simplicity. For an input
with no dependency on the past, we identify $u$ with $I$, corresponding
to $h(\tau)=\delta(\tau)$, which can be interpreted as $\tau_{m}$
(the membrane time constant) being much smaller than the time between
presentation of two consecutive inputs. This simplifying assumption
allows us to concentrate on the question we want to address here.
Note that neuron models formulated as leaky integrators with stochastic
spiking have been shown to perform well at reproducing and predicting
spike trains recorded in biological neurons \cite{Jolivet06,Jolivet2005}.
Note also that the approach developed here allows generalization to
the case of noisy inputs, as well as to the case of rate models, similar
to the formalism developed in \cite{Chechik}. The input current $I$
is the weighted sum of external inputs with activity $x$:

\begin{equation}
I=\sum_{j}V_{ij}x_{j}.
\end{equation}

\begin{figure}
\subfloat[]{ \includegraphics[scale=0.45]{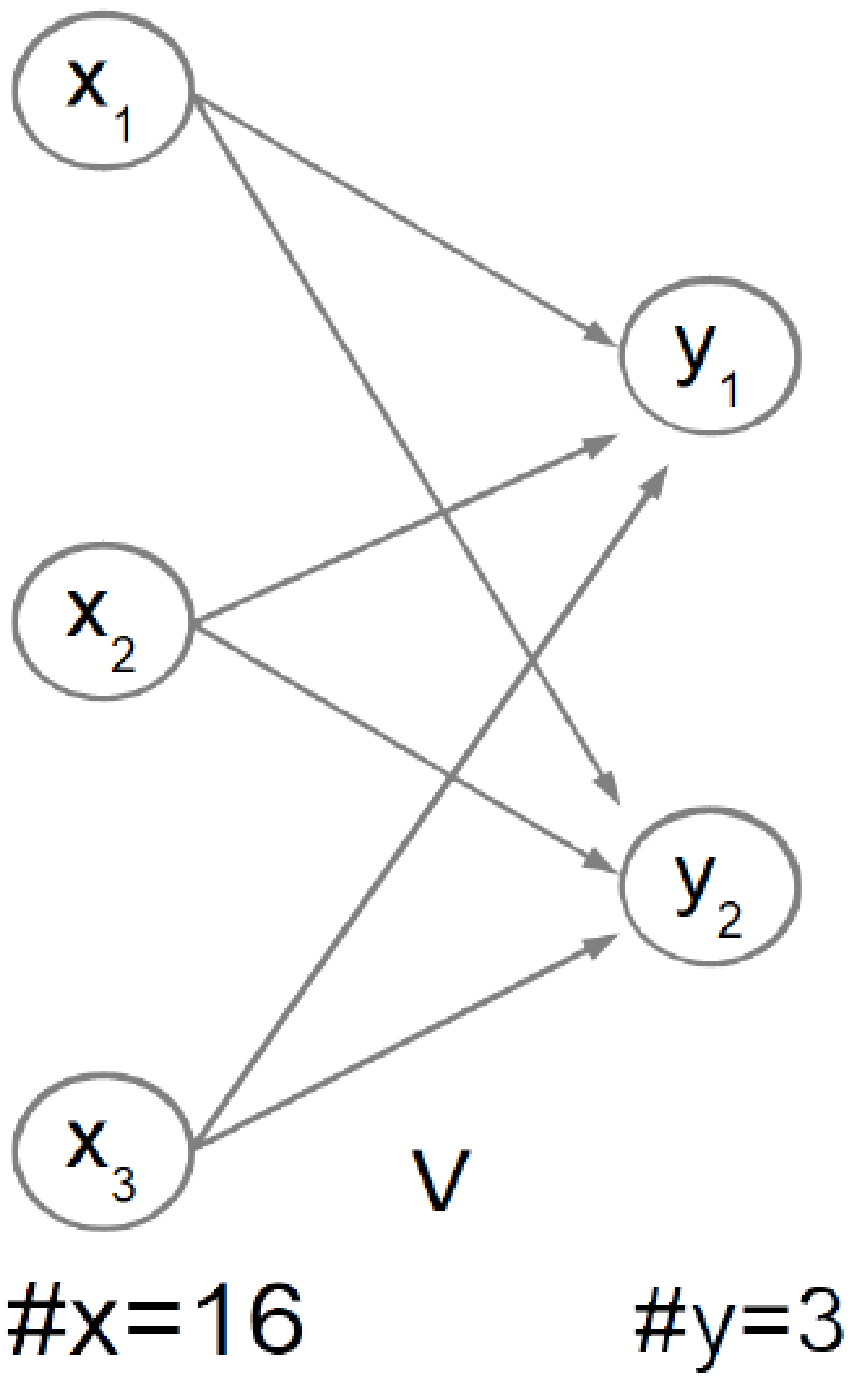}

}\subfloat[]{\includegraphics[scale=0.3]{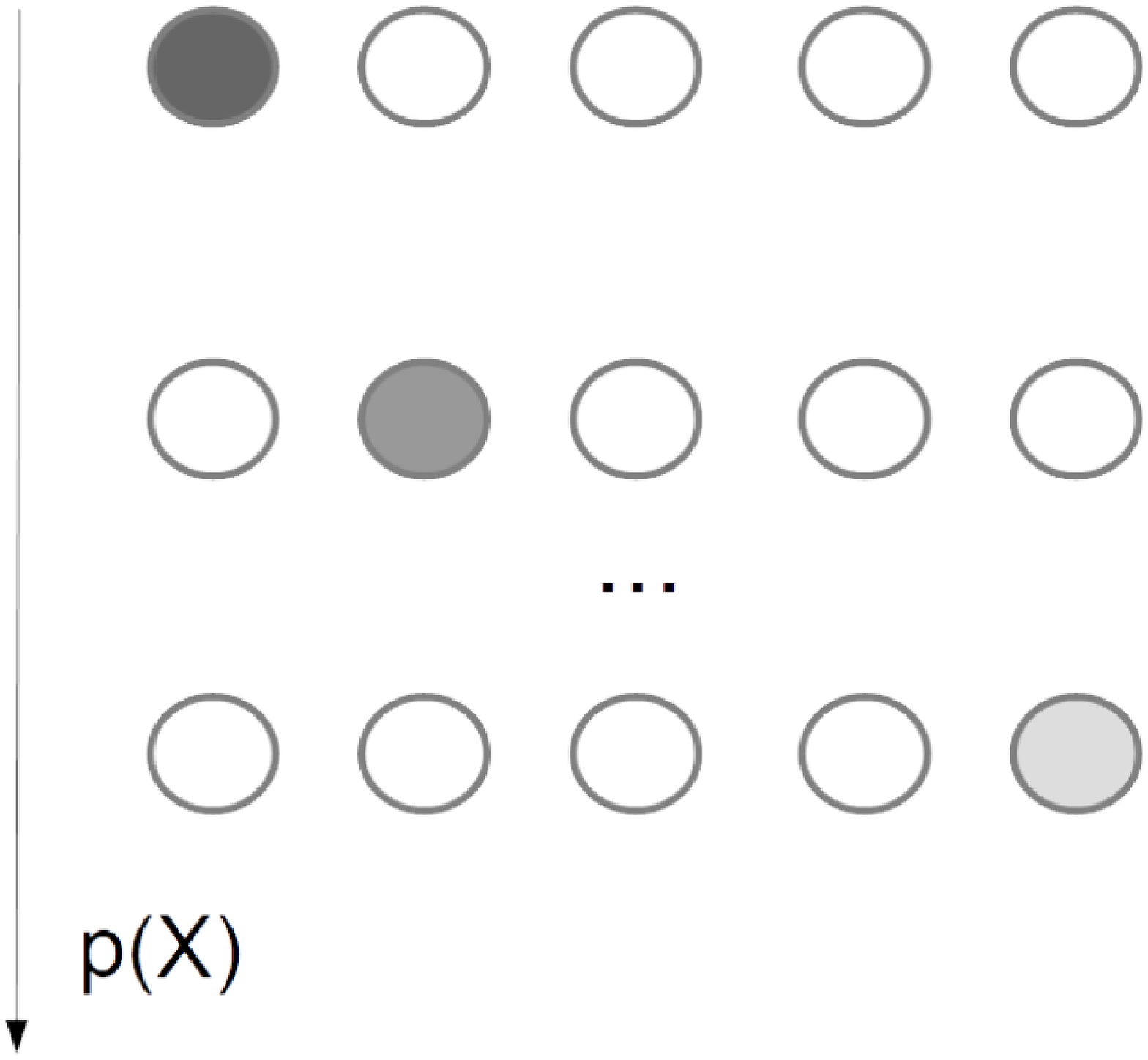}

}

\caption{\label{fig:1}(a) Basic scheme of the network considered here. An
array of inputs $x$ is connected via feed-forward connections $V$
to neurons generating outputs $y$. (b) Simple depiction of the inputs
applied in the simulations in Figures \ref{fig:2} and \ref{fig:3}.
Every input pattern contains only one (out of 16) active input channel.
The probability of a given $x$ to be applied to the network linearly
grows with the index number of this active channel (see text).}
\end{figure}

The weights of synaptic connections from input $j$ (with activity
$x_{j}$) to neuron $i$ (with activity $y_{i}$) are given by the
matrix entry $V_{ij}$. The network's architecture is schematically
shown in Figure \ref{fig:1}.

To obtain a set of learning rules, leading the network to optimization
of a desired function $F$ through learning, one can consider learning
as gradient descent optimization of $F$. Therefore, we put $\partial_{t}V_{ij}=\lambda\partial_{w_{ij}}F$
with some constant $\lambda$. For simplicity, we assume $\lambda=1$
and do not explicitely write it further. To quantitatively measure
information inference from the input $x$, we use mutual information:
\begin{equation}
M(x,y)=\sum_{x}p(x)\sum_{y}p(y|x)\ln p(y|x)-\sum_{y}p(y)\ln p(y)
\end{equation}
between inputs $x$ and outputs $y$. In order to balance information
inference and energy consumption, we now need to add an energy consumption
term $E$ to be subtracted from the information measure yielding:
\begin{equation}
F=M-\gamma E
\end{equation}
to be optimized, with $\gamma$ a parameter. In this case: 
\begin{equation}
\partial_{t}V_{ij}=\partial_{V_{ij}}F
\end{equation}
provides a learning rule balancing information representation and
energy consumption.

\section{Results}

\subsection{Learning rules derivation}

We first consider optimization of information inference with $F=M$.
Every neuron receives signals from input channels independent from
neuronal activity itself. This fact allows the decomposition $p(y|x)=\prod p_{i|x}$
with $p_{i}$ standing for the probability that the output neuron
$i$ spikes ($y_{i}=1$), or doesn't spike ($y_{i}=0$). $p(y_{i}=1|x)=f(I_{i})$
with $I_{i}=\sum_{j}V_{ij}x_{j}$ as the total input to the neuron
$i$. Conversely, $p(y_{i}=0|x)=1-p(y_{i}=1|x)$. One can then write
$p_{i}=y_{i}f(I_{i})+(1-y_{i})(1-f(I_{i}))$. Using that, we get the
expression for $\partial_{V_{ij}}M$: 
\begin{equation}
\partial_{V_{ij}}M=\sum_{x}\sum_{y}p(x)p(y|x)/p_{i|x}x_{j}(2y_{i}-1)f^{/}(I_{i})\,(\ln p(y|x)-\ln p(y)),
\end{equation}
which with $p(x)p(y|x)=p(x,y)$ results in the rule for $V$ updates:
\begin{equation}
\Delta V_{ij}=x_{j}(2y_{i}-1)f^{/}(I_{i})/\{y_{i}f(I_{i})+(1-y_{i})(1-f(I_{i})\}\,(\ln p(y|x)-\ln p(y))
\end{equation}
with $\ln p(y|x)=\sum_{i}\ln p_{i|x}=\sum_{i}\ln\{y_{i}f(I_{i})+(1-y_{i})(1-f(I_{i}))\}$
for every time an input is presented. $p(y_{i})$ can be expressed
via the averaged activity $\bar{y}_{i}=<y_{i}p(y_{i})>=p(y_{i}=1)$,
and for a system of only one output neuron, or in the approximation
of uncorrelated output neurons, $\ln p(y)=\sum_{i}\ln(y_{i}\bar{y_{i}}+(1-y_{i})(1-\bar{y_{i}}))$.
For these cases, the learning rule can be formulated explicitly after
summing over $y_{i}\in\{0,1\}$ as: 
\begin{align}
\partial_{t}V_{ij} & =\partial_{V_{ij}}M=\sum_{x}p(x)xf^{/}(I)\bigg\{\ln\frac{p(y=1|x)}{p(y=0|x)}-\ln\frac{p(y=1)}{p(y=0)}\bigg\}\nonumber \\
 & =\bigg\langle xf^{/}(I)\bigg\{\ln\frac{f(I)}{1-f(I)}-\ln\frac{\bar{y}}{1-\bar{y}}\bigg\}\bigg\rangle=<xf^{/}(I)(g(f(I))-g(\bar{y})>
\end{align}
with $g(\epsilon)=\ln\frac{\epsilon}{1-\epsilon}$. This rule is remarkably
similar to the Bienenstock-Cooper-Munro (BCM) learning rule (it differs
in the particular form of the function $g$) \cite{Intrator92}.

Usually, however, the outputs caused by shared inputs are not independent
from each other, and for a large output layer, a direct estimation
of $p(y)$ is not obvious. One of a number of natural suggestions,
which we used in simulations, is to approximate $p(y)$ using probabilities
of states of single neurons, and of pairs of single neurons, which
can be obtained by measuring the output covariance $<y_{i}y_{j}>$
of neuron pairs. Assuming vanishing non-trivial third and further
cumulants, $p(y)$ can be written as a sum of products of $p(y_{i},y_{j})$.
One of the possible simple implementable approximations is $\ln p(y)\sim\sum_{i}\ln p(y_{i})+\ln\frac{1}{N(N-1)}\sum_{i,j\neq i}\frac{p(y_{i},y_{j})}{p(y_{i})p(y_{j})}$.

\subsection{Energy supply}

The computational power of a neural network is bound by the amount
of available resources, such as a restricted energy supply. To take
into account a desirable minimization of the energetic costs of computation,
we introduce the energy term: 
\begin{equation}
E=\bigg\langle\sum_{i}(c_{sp}y_{i}+c_{psp}\sum_{j}V_{ij}x_{j})\bigg\rangle=\sum_{x}p(x)\bigg(\sum_{y}\{(\sum_{i}y_{i})p(y|x)\}+c_{psp}\sum_{j}V_{ij}x_{j}\bigg)
\end{equation}
and optimize $\alpha M-\beta E$, or, for shortness, $F=M-\gamma E$
with $\gamma=\frac{\beta}{\alpha}$, with $\alpha$ and $\beta$ reflecting
the relative importance of information inference and energy consumption.
The first term in the expression for $E$ describes the energetic
costs of output spikes (with constant $c_{sp}$), and the second the
energetic costs of postsynaptic potentials (with constant $c_{psp}$).
$\partial_{t}V_{ij}=\partial_{V_{ij}}F=\partial_{V_{ij}}M-\gamma\partial_{V_{ij}}E$
gives the new learning rule optimizing $F$. $\partial_{V_{ij}}M$
was obtained in the previous subsection, and thus: 
\begin{align}
\partial_{V_{ij}}E & =\sum_{x}p(x)\bigg\{ c_{sp}\sum_{y}\{\partial_{V_{ij}}p(y|x)(\sum_{k}y_{k})\}+c_{psp}x_{j}\bigg\}\nonumber \\
 & =\sum_{x}p(x)\bigg\{ c_{sp}\sum_{y}\bigg\{ p(y|x)\frac{\partial_{V_{ij}}p(y_{i}|x)}{p(y_{i}|x)}(\sum_{k}y_{k})\bigg\}+c_{psp}x_{j}\bigg\}\nonumber \\
 & =\sum_{x}p(x)\{c_{sp}x_{j}f^{/}(I_{i})+c_{psp}x_{j}\}\nonumber \\
 & =\sum_{x}p(x)x_{j}\bigg\{ c_{sp}p(y_{i}=1|x)\frac{f_{i}^{/}(I_{i})}{f(I_{i})}+c_{psp}\bigg\},
\end{align}
where we grouped possible values of $y$ in pairs differing only in
values of $y_{i}$. The resulting learning rule is given by: 
\begin{equation}
\partial_{t}V_{ij}=\sum_{x}\sum_{y}p(x)\bigg\{\frac{p(y|x)}{p_{i|x}}x_{j}(2y_{i}-1)f^{/}(I_{i})\,(\ln p(y|x)-\ln p(y)-\gamma c_{sp})-\gamma c_{psp}\bigg\},
\end{equation}
and implementing weight updates every time an input is presented:
\begin{equation}
\Delta V_{ij}=x_{j}f^{/}(I_{i})(2y_{i}-1)/p_{i|x}\bigg\{(\ln p(y|x)-\ln p(y))-\gamma c_{sp}(\sum_{i}y_{i})\bigg\}-\gamma c_{psp}x_{j}.\label{eq:dv_E}
\end{equation}

Alternatively, we get $\Delta V_{ij}=x_{j}f^{/}(I_{i})\,\{(2y_{i}-1)(\ln p(y|x)-\ln p(y))/p_{i|x}-\gamma c_{sp}\}-\gamma c_{psp}x_{j}$
or $\Delta V_{ij}=x_{j}f^{/}(I_{i})(2y_{i}-1)/p_{i|x}\,\{(\ln p(y|x)-\ln p(y))\}-\gamma c_{sp}y_{i}\}-\gamma c_{psp}x_{j}$.
Different implementations of the rule for weight updates are possible
because of the relation between $f(I_{i})$ and $p(y_{i}=1)$. However,
for slow enough learning, all of them produce the same weights dynamics,
and since the last two forms do not require additional summation over
all neurons to obtain the total current activity of the network, they
make for a simpler implementation. The first form yields simpler understanding
of the weights' evolution tendencies and was used in simulations.

\begin{figure}
\subfloat[]{\includegraphics[scale=0.3]{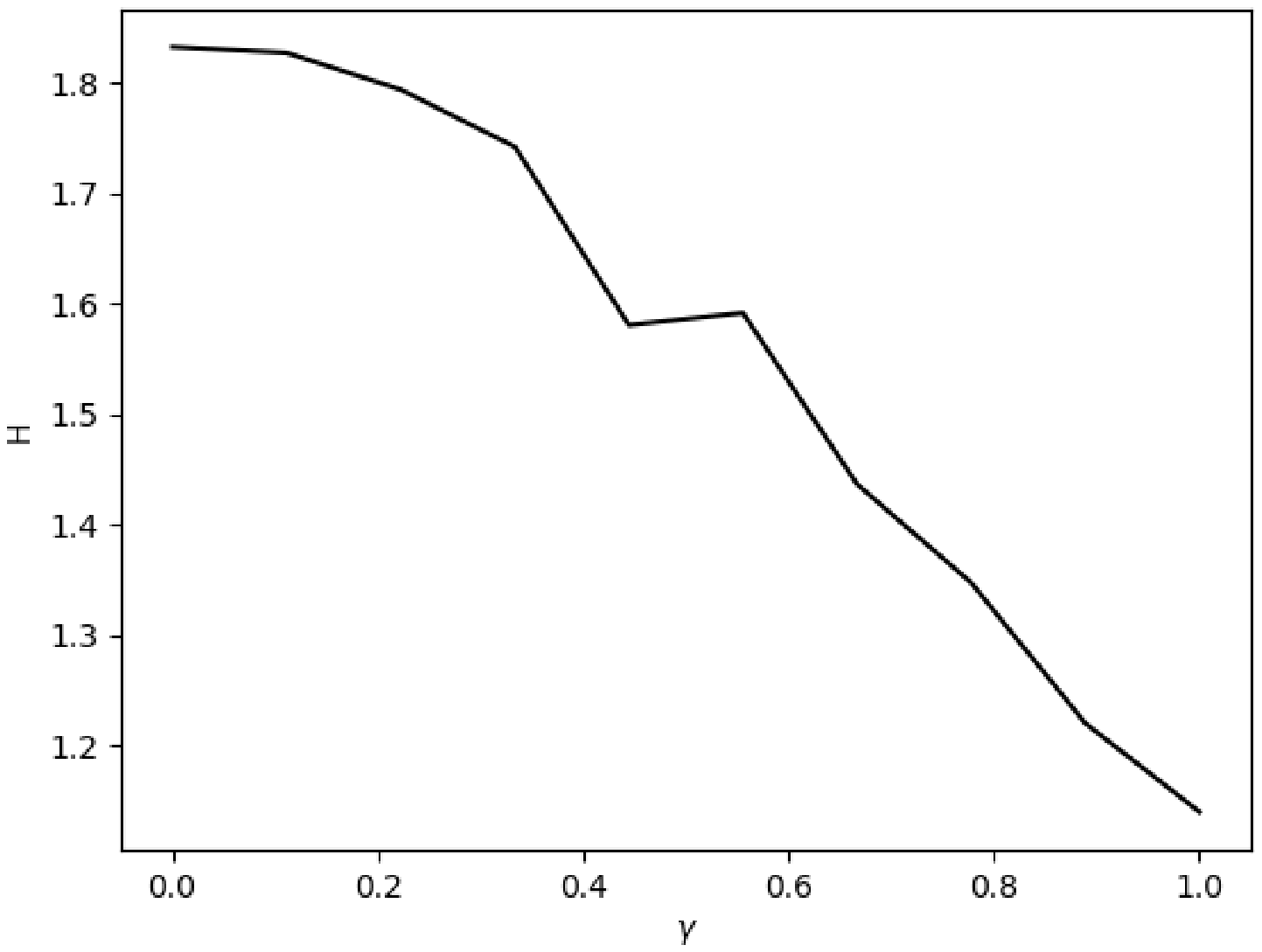}

}\subfloat[]{\includegraphics[scale=0.3]{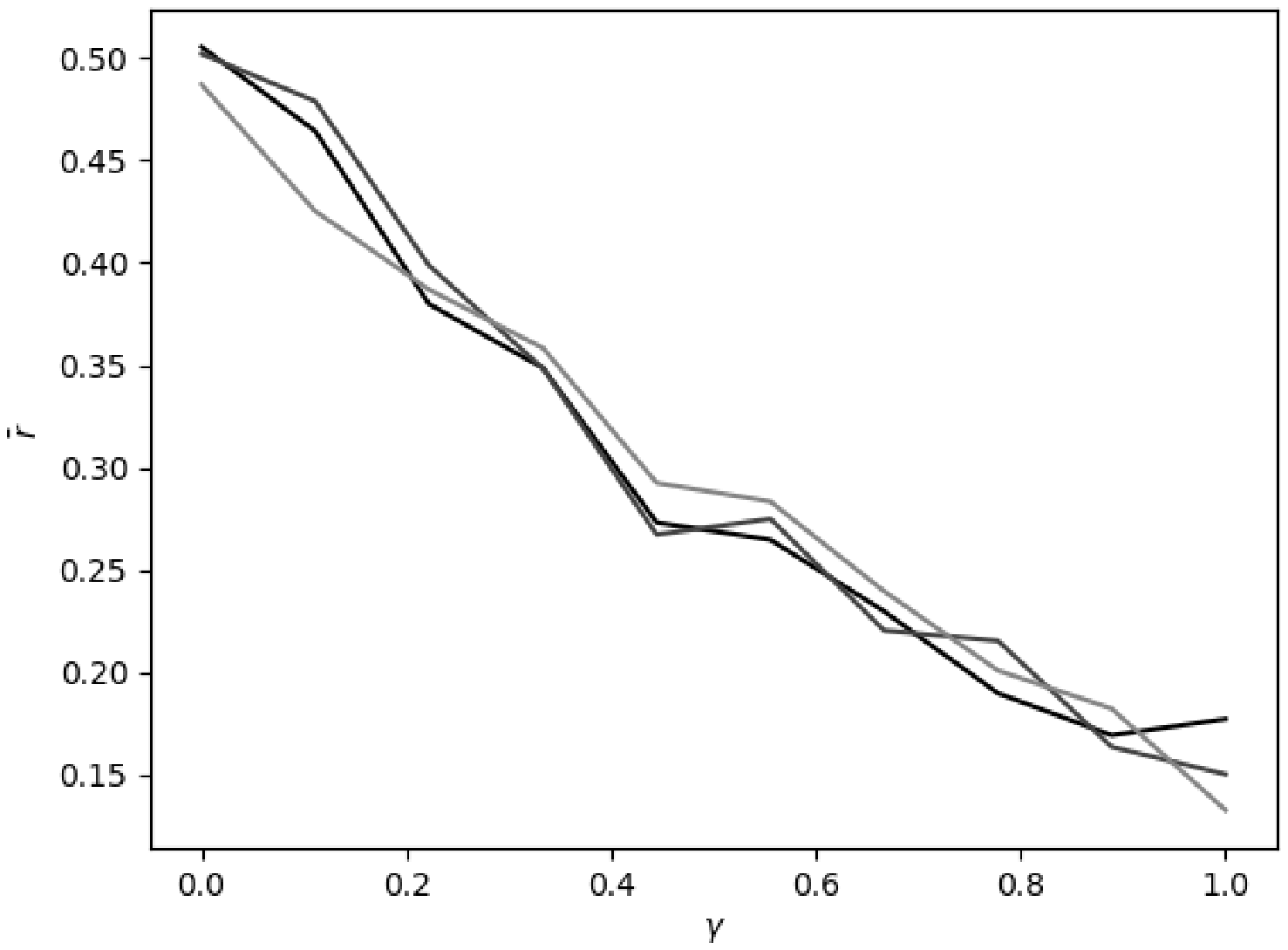}

}

\subfloat[]{ \includegraphics[scale=0.3]{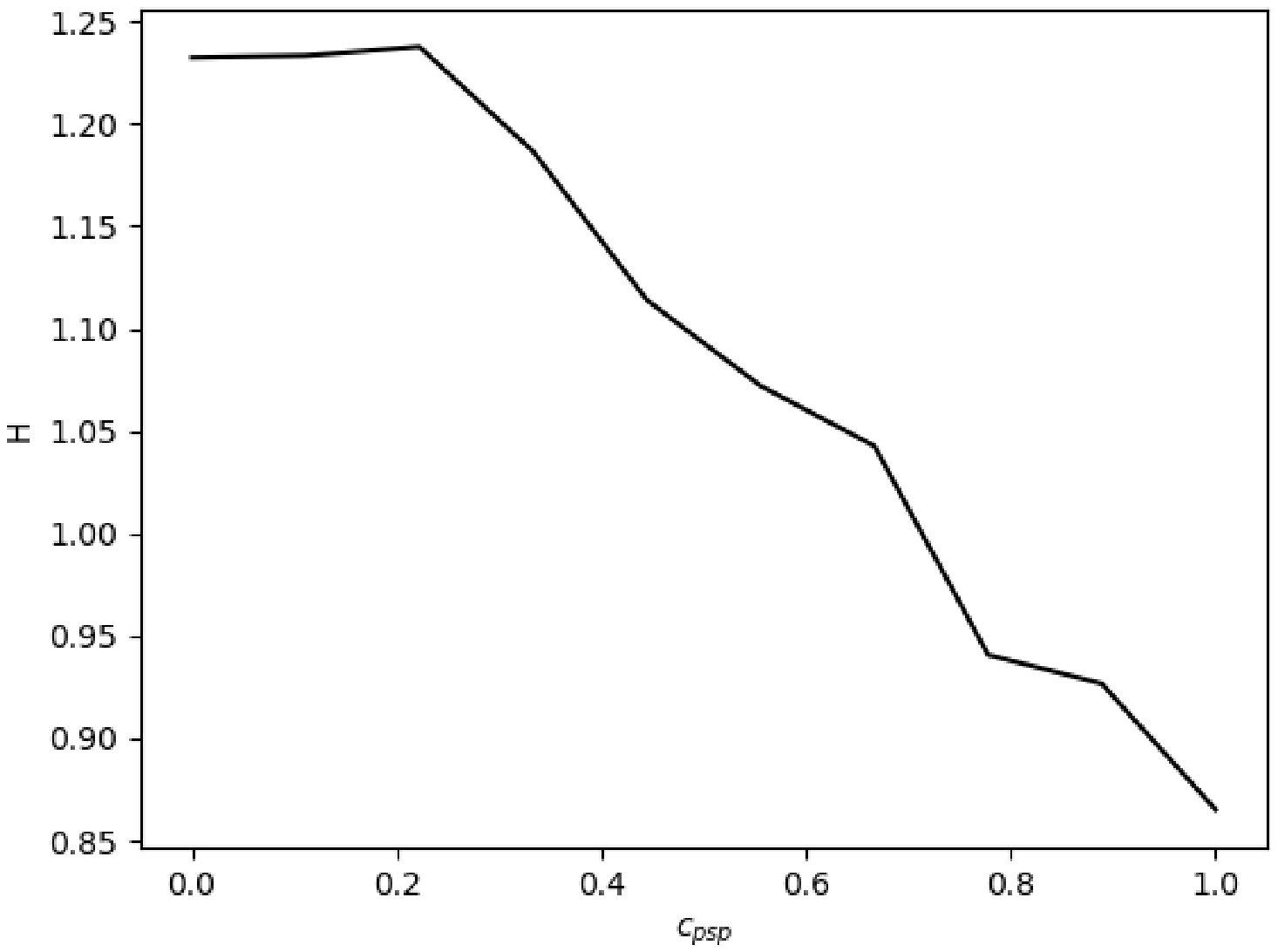}

}\subfloat[]{\includegraphics[scale=0.3]{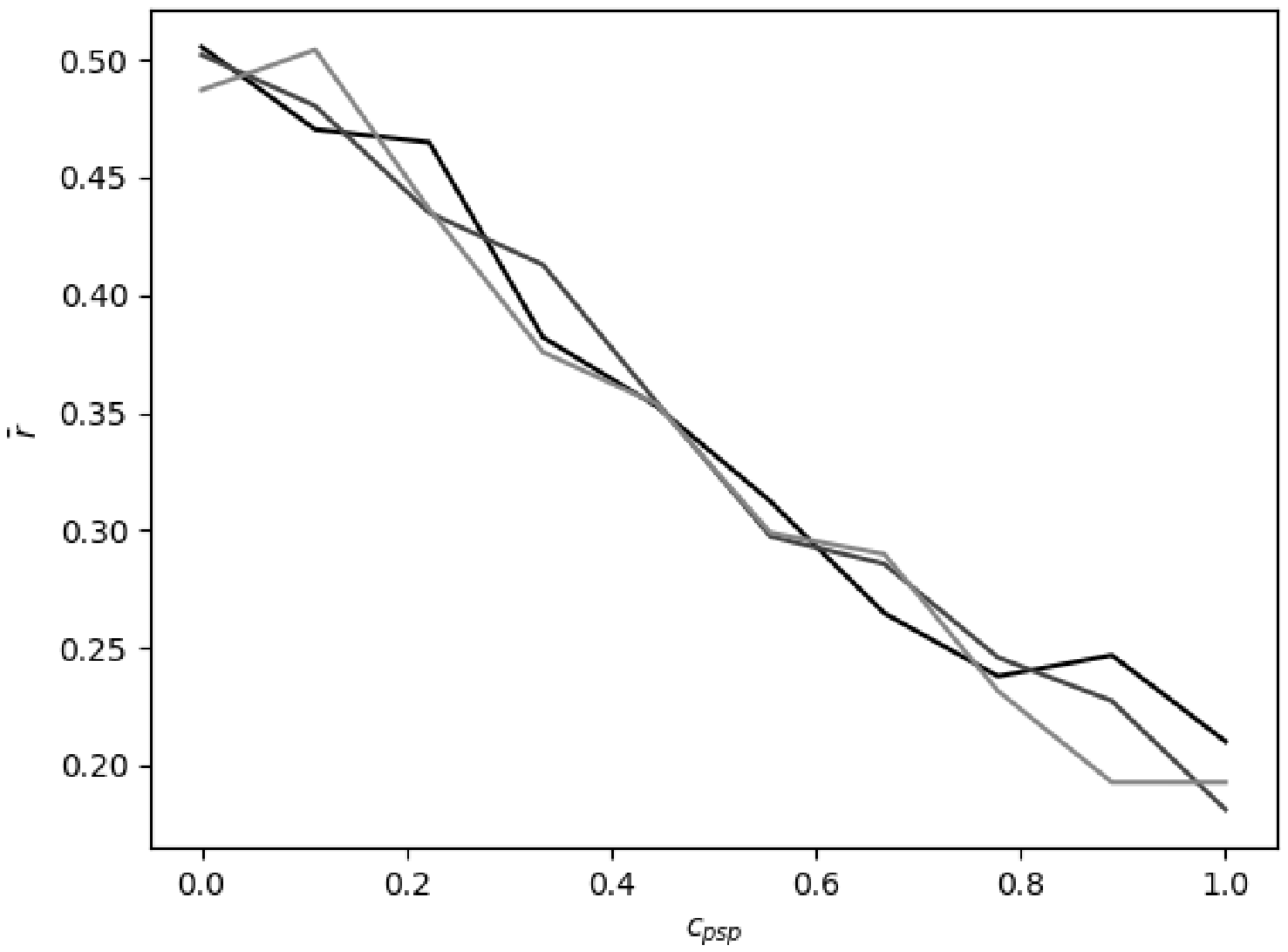}

}

\caption{\label{fig:2}(a) Mutual information between the network's inputs
and outputs decreases as energetic constraints are applied ($\gamma>0$).
(b) Similarly, the activity of the $N=3$ output neurons decreases
as energetic constraints are applied. In both (a) and (b), $c_{sp}=1$
and $c_{psp}=0$. (c) Same as in (a) but plotted versus the energetic
cost associated with post-synaptic potentials ($c_{psp}$). (d) Same
as in (b) but plotted versus the energetic cost associated with post-synaptic
potentials. In both (c) and (d), $\gamma=1$ and $c_{sp}=0$.}
\end{figure}

The terms proportional to $\gamma$ contribute to a reduction in precision
of information inference to limit energy consumption. This is illustated
in simulations for a simple network with $N=3$ output neurons and
$K=16$ input neurons (Figure \ref{fig:2}). Every presented input
contains only one active input channel, with the probability of appearence
of the input with the $k$'th active channel being proportional to
$k\in[1,...,K]$ $p(x_{k})=(k_{0}+k)/s_{k}$ with normalization factor
$s_{k}=\frac{K(K+1+2k_{0})}{2}$. For simulations, $f(u)=1/(1+\exp(-b(u-u_{0})))$
with $u_{0}=0.5$ and $b=10$, and $k_{0}=15$.

\begin{figure}
\subfloat[]{\includegraphics[scale=0.3]{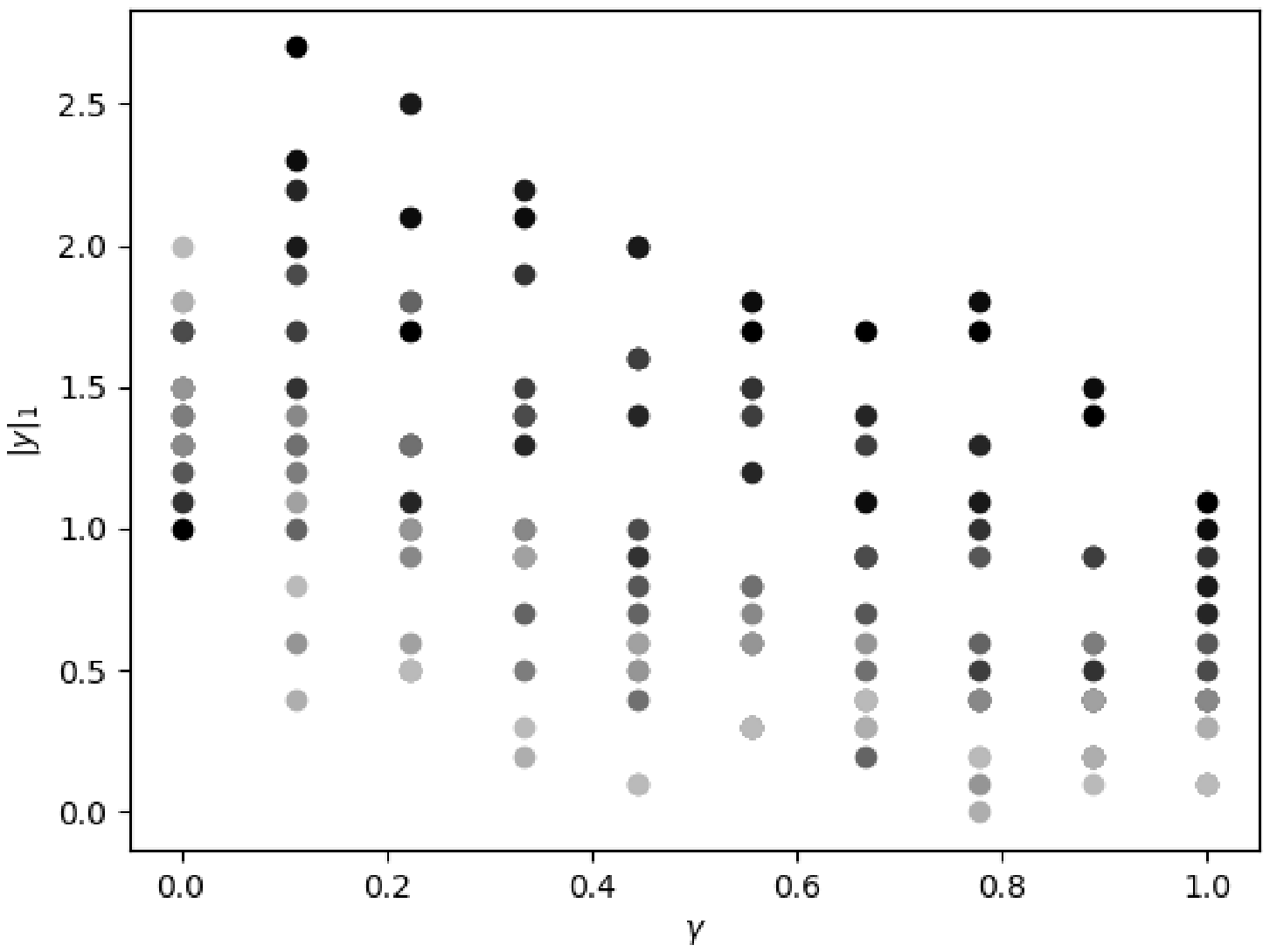}

}\subfloat[]{\includegraphics[scale=0.3]{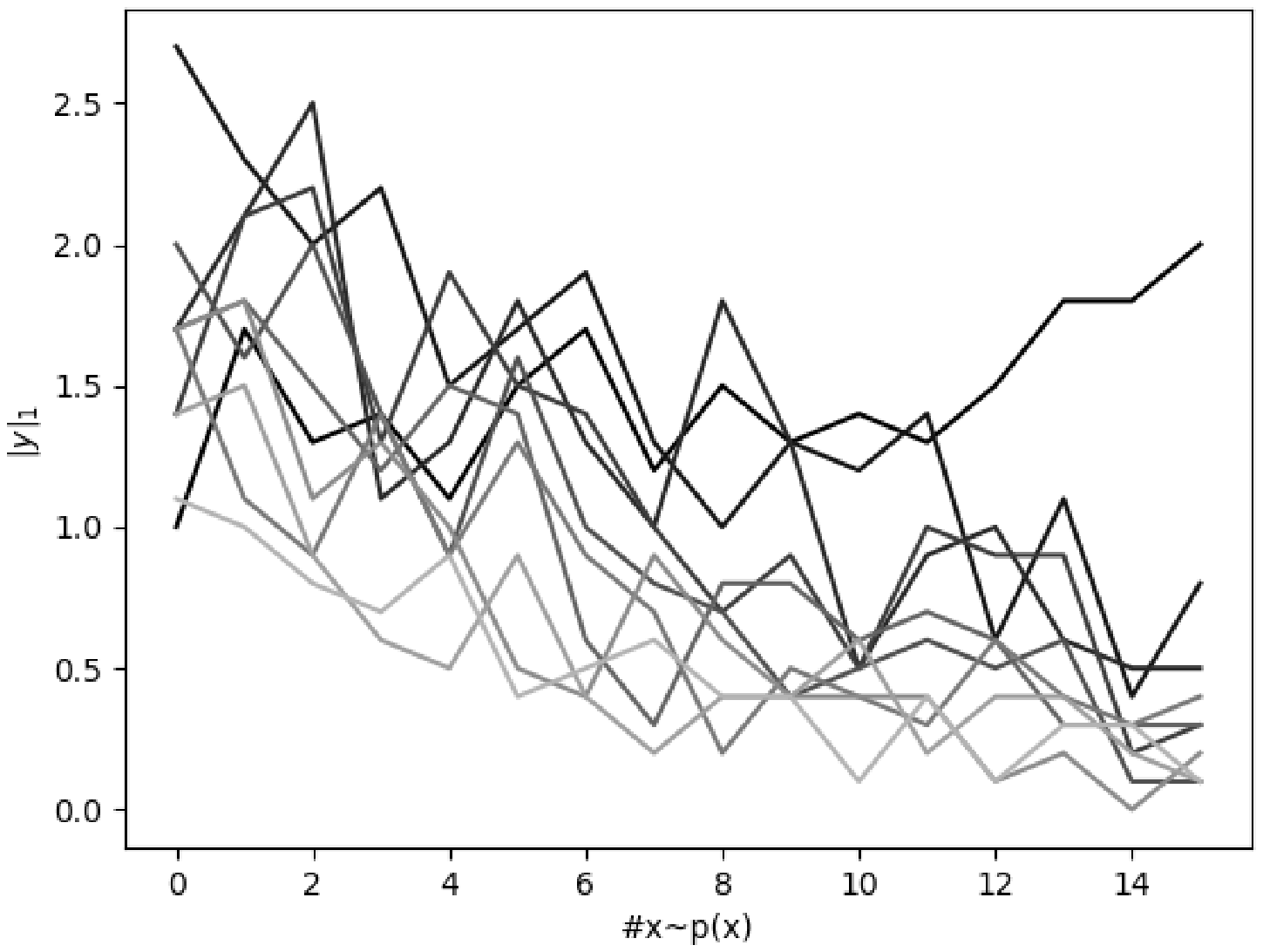}

}

\subfloat[]{\includegraphics[scale=0.3]{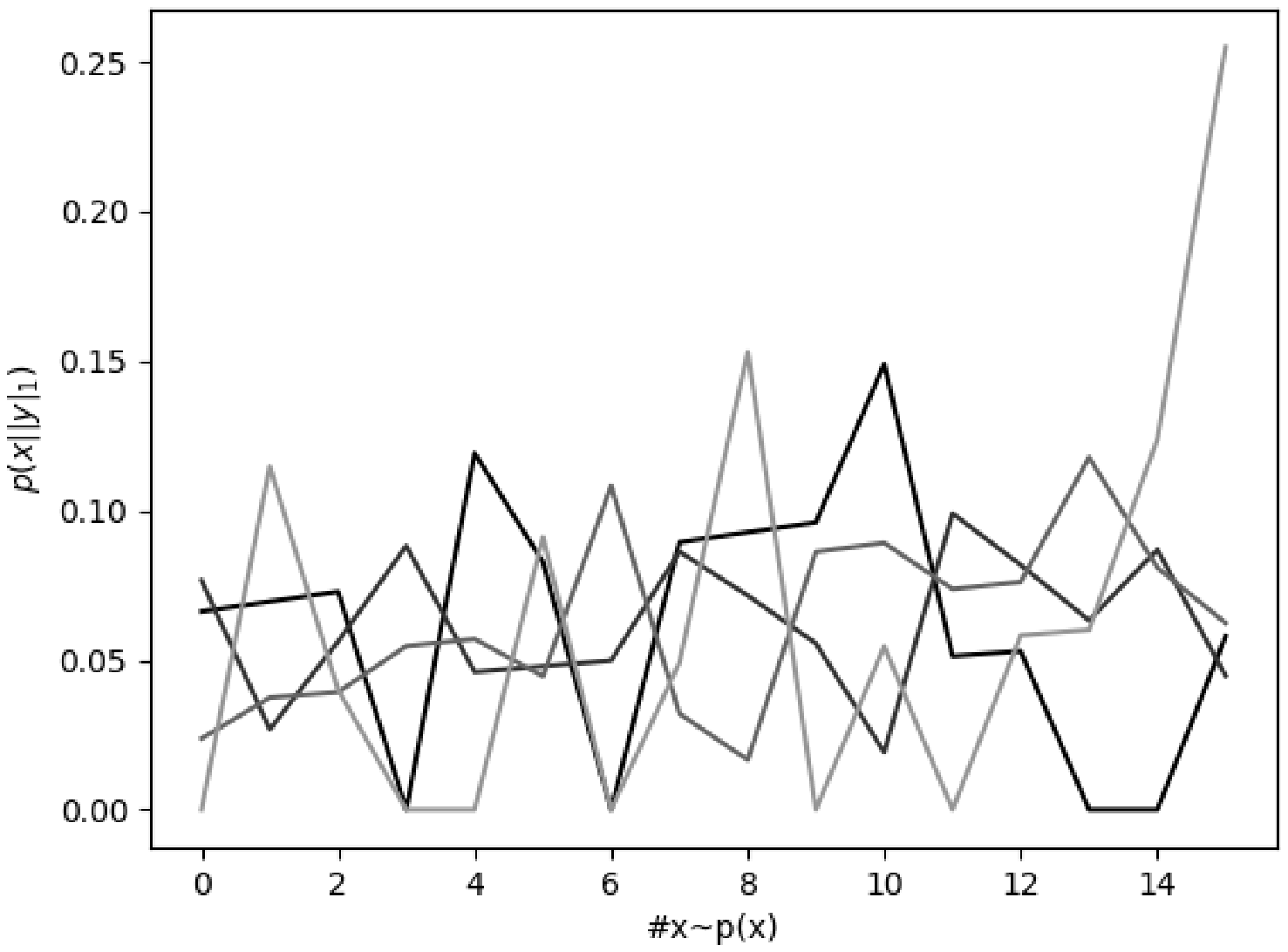}

}\subfloat[]{ \includegraphics[scale=0.3]{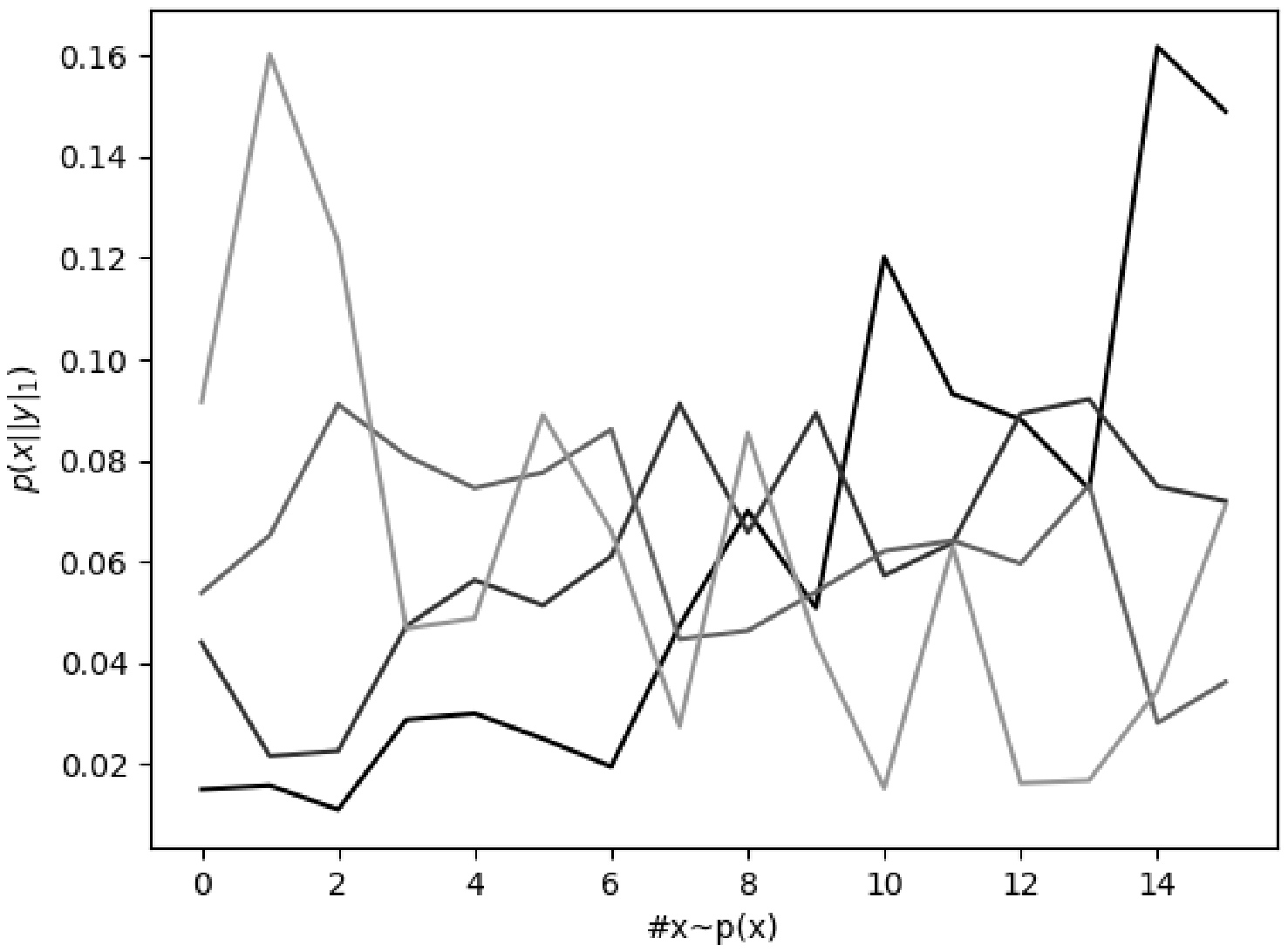}

}

\caption{ \label{fig:3}Energetic constraints map frequent inputs to low energy
orbits. (a) Averaged network activity $|y|_{1}$ in response to inputs
with increasing probabilities $p_{x}$ plotted versus the energetic
cost of spikes (with $c_{sp}=1$ and $c_{psp}=0$). The probability
of inputs is coded in shades of gray, from least (black) to most probable
(increasingly lighter shades of gray). When energetic constraints
are absent ($\gamma=0$), inputs presented more frequently evoke larger
responses in the network and there is a distinct correlation between
frequency of presentation and intensity of evoked responses (dots
roughly organised from most probable, light gray, on top, to least
probable, black, at the bottom). This relation starts to invert as
$\gamma>0$ and is completely reverted at $\gamma=1$, with the most
probable inputs (light gray) now being mapped onto orbits with very
low averaged network activity, and the least probable inputs (black)
now being mapped onto orbits with higher activity. (b) Averaged network
activity $|y|_{1}$ evoked by inputs in different channels for $\gamma=0$
(black) and increasing values of $\gamma>0$ (increasinlgy lighter
shades of grey). At $\gamma=0$, evoked activity is more or less even
whether inputs are presented with low or high probability. As $\gamma$
increases, inputs presented with high probability (channels with high
indices) evoke low activity while inputs presented with low probability
(channels with low indices) evoke high activity. The overall activity
decreases as $\gamma$ increases (from dark to light shades of grey).
(c) and (d) illustrate the mapping of inputs induced by energetic
contraints. (c) Probability of presentation of a specific input channel
given the observed summed output activity $|y|_{1}=0$ (black) or
$|y|_{1}=1,2,3$ (increasingly lighter shades of grey). In the absence
of energetic constraints ($\gamma c_{sp}=0$), all input channels
are more or less equiprobable. (d) Same as in (c) but with energetic
constraints ($\gamma c_{sp}=0.3$). Now the most probable inputs (\#14
and \#15) most likely evoke no activity (black line), while rare inputs
(\#0, \#1 and \#2) most likely evoke maximal activity (all $N=3$
output neurons activated; lightest line).}
\end{figure}

Adding energetic considerations to the derivation of the learning
rule has the effect of rearrenging the correspondence between inputs
and outputs. For highly probable outputs, $-\ln p(y)$ has low values.
So, at equilibrium, more energetically expensive states with a high
number of active neurons are more rare. On the other side, at the
onset of learning, the term $\{(\ln p(y|x)-\ln p(y)-\gamma(\sum_{i}y_{i})\}$
is more often positive for energetically cheap network responses.
As a consequence, input patterns occuring with high probabilities
will evoke energetically cheap responses, involving a low number of
neurons into coding. Rare events, on the contrary, can recruit a higher
number of neurons for coding. Simulation results illustrating this
rearrengement are presented in Figure \ref{fig:3}, showing the relation
between $|y|_{1}=\sum_{i}y_{i}$ and $p(x)$, as well as with $\gamma$.
One way to interpret these results is to see every input $x$ as searching
during learning for some output state $y$ to occupy, with $y$ states
already occupied by a different input being more difficult to occupy.
This effect is compounded by energetic terms. Additionally, rare events
only lead to small modifications in the formation of connections shared
with inputs presented more often. In the next subsection, we consider
unreliable synapses and demonstrate that in this case too, rare inputs
are effectively cut off by learning.

\subsection{Inference with unreliable input channels}

Input signals reaching the brain from primary sensory neurons can
also be unreliable. This fact can be modelled by adding Gaussian noise
with variance $\sigma_{nz}$ to the input channels. In this scenario,
large synaptic weights will amplify the noise and should be avoided.
Derivation of learning rules taking input noise into account leads
to an additional term proportional to $-\sigma_{nz}^{2}V$, which
effectively implements a form of regularization. Similar considerations
are also presented in reference \cite{varlearn}. The derivation is
given in the Appendix.

\begin{figure}
\subfloat[]{\includegraphics[scale=0.3]{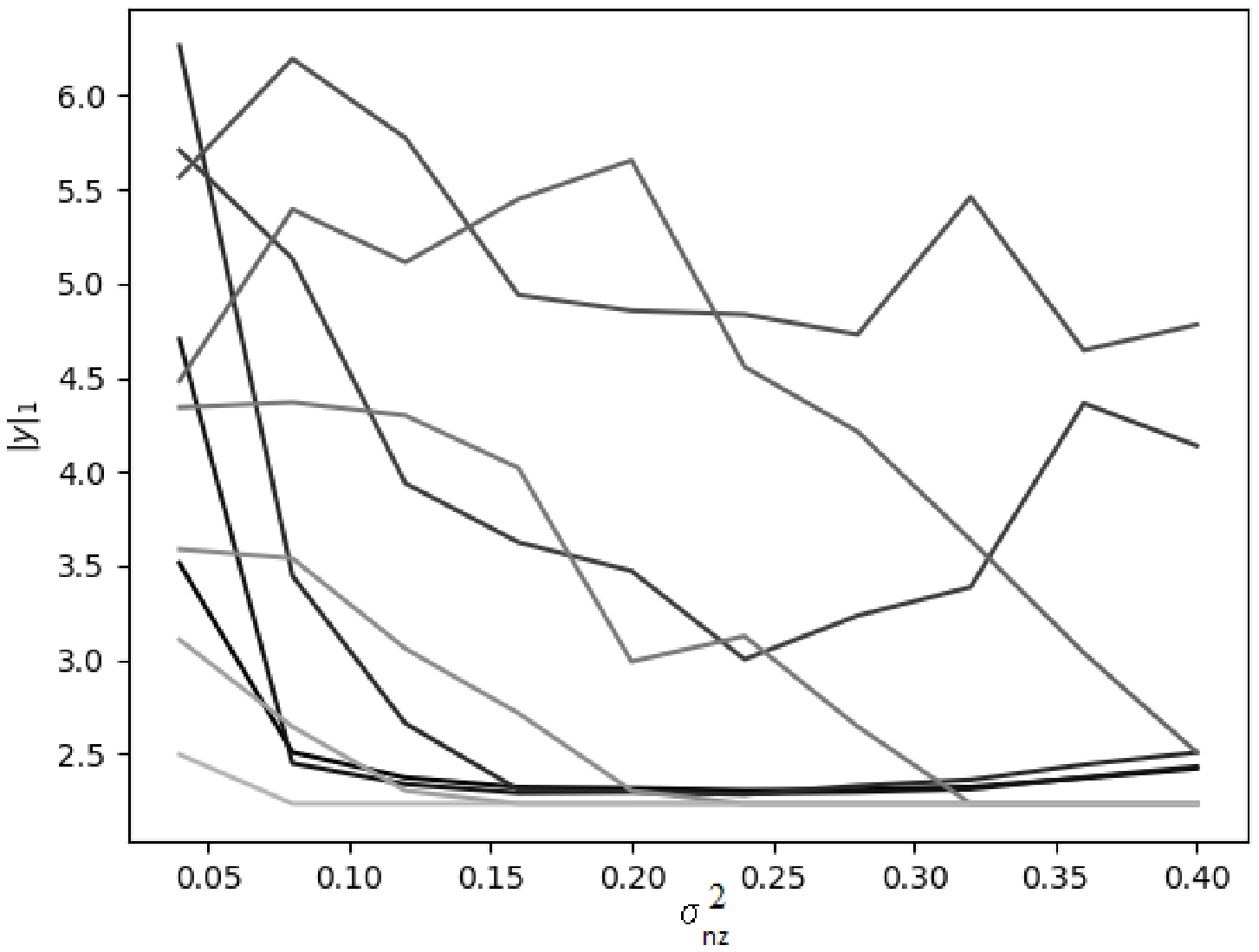}

}\subfloat[]{\includegraphics[scale=0.3]{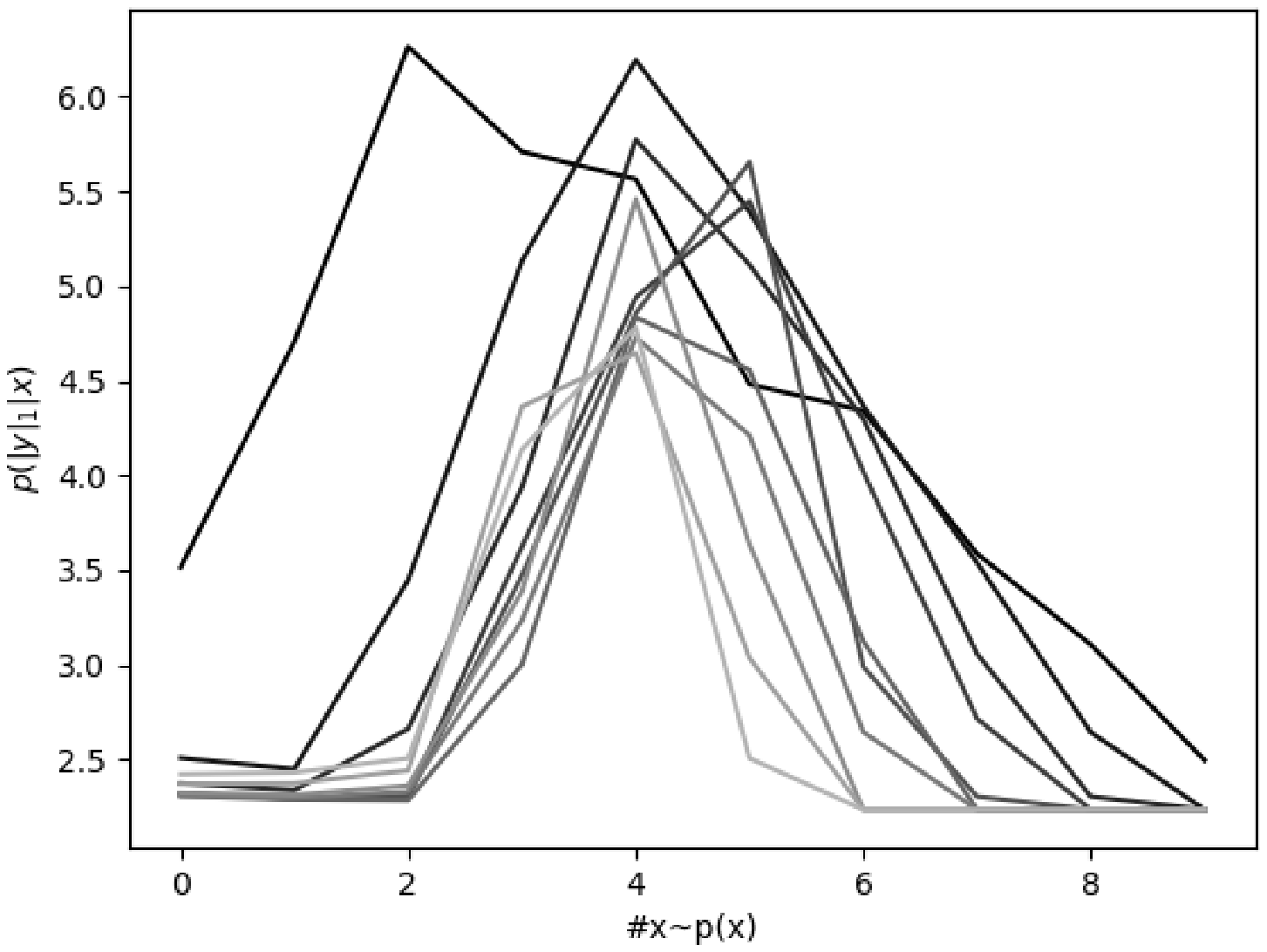}

}

\caption{\label{fig:4}Learning in the presence of noise. (a) Averaged activity
$|y|_{1}$ of the network's response with $\gamma=1$, $c_{psp}=0.1$,
$c_{sp}=0.3$ and input noise $\sigma_{nz}^{2}$ changing in arithmetic
progression from $0.04$ to $0.4$ for different input patterns from
low probability of occurence (black line) to higher probabilities
of occurence (increasingly lighter shades of grey). As the noise level
increases, inputs with low probability of occurence (darker shades)
exhibit decaying activities and, eventually, aren't learned by the
network. (b) Probability of presentation of a specific input channel
given the observed summed output activity $|y|_{1}$ is plotted for
increasing noise intensities $\sigma_{nz}^{2}$ (dark to light shades
of grey). As in Figure 2b, activity decays with the probability of
presentation (increasing channel index). Additionally, the presence
of noise ($\sigma_{nz}^{2}>0$) leads to a reduction of activity for
channels corresponding to very rare inputs that cannot be distinguished
from the noise. As a result, the peak of activity occurs for the same
moderately probable $p_{x}$. The width of the distribution decreases
with increasing noise intensities.}
\end{figure}

This regularization term prevents the network from learning rare input
events, whose low appearance frequency does not allow distinguishing
them from random inputs induced by the noise in input channels. As
a result, the evoked network responses grow in intensity as the probability
of input patterns decay, until said probability decays down to the
noise level. For input probabilities beyond that, and below the noise
level, the network's activity vanishes. As a consequence, the network's
response is strongest for moderately probable input patterns, rare,
but still distinguishable from the noise. This is illustrated in Figure
\ref{fig:4}.

\section{Conclusions and Discussion}

The human brain consumes vasts amounts of energy with respect to its
mass in the body (see \cite{Jolivet12} for a recent review). It is
thus natural to assume that energetic constraints play an important
role in shaping activity patterns and learning in neural networks,
and a large body of experimental literature supports this assumption.
Here, we addressed this question in a simple neural network model
consisting of stochastic non-linear Hawkes neurons. Specifically,
we started by deriving for this neural network a simple learning rule
maximising information inference between inputs and outputs. For this,
we used mutual information as the measure of information inference
quality.

That learning rule includes terms of temporally and spatially local
covariances, and a non-local value common to all neurons. A similar
structure for learning rules, also derived by gradient descent on
a goal function, is reported in \cite{ICA}, where the learning rule
is derived as optimization of the squarred error for the network performing
independent component analysis, and in \cite{Brea}, optimizing Kullback-Leibler
divergence by teaching the network to reproduce a given set of input
patterns developing in time. The non-local term is a function of the
whole network activity and can be interpreted, e.g., as a slow non-local
signal distributed throughout the network such as a neuromodulator
\cite{chem2,chem3,chem4,chem5}, GABA \cite{chem1}, $NO$ \cite{NO},
or maybe a glial signal \cite{chem6}. The local covariance-sensitive
parts of the learning rule could be realized via spike-time dependent
plasticity mechanisms. The difference between those terms \textendash{}
local temporal and spatial covariances on one hand, and the non-local
value averaged over time on the other hand \textendash{} implies a
plasticity mechanism comparing the properties of ongoing activity
to its recent past, similar to the BCM learning rule, which is also
known to be related to optimization of information measures \cite{Intrator92}.
Qualitatively, terms in $M$ including $p(y|x)$ characterize the
specificity of the network's responses to various inputs, while terms
including $p(y)$ characterize the efficient use by the network of
the output space. This was implemented here with the covariance of
single pairs of neurons, each of which can be measured locally. Another
possible implementation could be the existence of intermediate inhibitory
neurons in the network in a manner described in \cite{Maass13}, but
this will require further studies.

Taking energy consumption minimization into account results in new
effects due to the additional terms in the learning rule. First, mutual
information reaches lower saturation values, representing the trade-off
between the quality of information inference and smaller synaptic
weights, which limit energy consumption. Also, the probability of
energetically expensive network responses decreases, and these become
less probable than energetically cheaper responses. Regularly occuring
inputs tend to evoke energetically cheap responses involving fewer
neurons. Finally, taking input noise into account modifies learning
so that very rare inputs are ignored, if they are so rare that they
can't be distinguished from the noise. If the number of neurons in
the network is large enough to allow for a unique representation of
every input, i.e. $p(y|x)$ is approximately 1 or 0, then from Equation
(\ref{eq:dv_E}), it follows that $p(y)\sim\exp(-\gamma c_{sp}|y|)$
with $|y|=\sum_{i}y_{i}$ for $c_{psp}=0$, with $c_{psp}>0$ only
reducing $p(y)$. A similar exponential dependency is obtained in
\cite{Tsubo} when the network optimizes information representation
under energy constraints for the probability of a random neuron to
exhibit a given rate value. Both results can be seen as a Huffman-like
\cite{Huffman} economical coding scheme, whereby energy savings are
also taken into account.

In further studies, we plan to apply the method demonstrated here
to time-dependent inputs, networks of excitatory and inhibitory neurons,
and population coding. Similar to the ideas applied in \cite{Brea}
and \cite{Maass14} for likelihood measures, we propose to apply the
approach presented here to time-dependent inputs considering every
time sequence of a given time length as a single input, and optimizing
a goal function for a set of these inputs. We have not yet taken into
account Dale's principle, the experimentally observed separation of
neurons into distinct excitatory and inhibitory populations with exclusively
positive, or negative, outgoing weights. Preliminary results for time-dependent
inputs suggest a rough connection between inhibition and energy savings.
Although generation of inhibition also requires energy (but see \cite{chatton2003}),
inhibitory effects help to avoid non-necessary spiking. In agreement
with this thought, oscillations can also be considered as a self-organized
way of saving energy by population coding, as recently demonstrated
in \cite{Gutkin_osc}. More generally, one can study the relation
between excitation, inhibition, the competition between information
inference and energy savings, and self-organized criticality in neural
networks inspired by arguments recently reviewed in \cite{EIrev}.

\section*{Appendix}

If neurons receive noisy inputs $x+\xi$ with $<\xi>=0,\ <\xi^{2}>=\sigma_{nz}^{2}$
when an input $x$ is presented, then the probability to fire in response
to input $x$ is $p(y=1|x)=\int f(V(x+\xi))d\xi=G(Vx,|V\sigma_{nz}|^{2})$
with $G(I,\sigma_{I}^{2})$ defined as $G=\frac{1}{\sqrt{2\pi}\sigma_{I}}\int f(I+\xi)\exp(-\frac{\xi^{2}}{2\sigma_{I}^{2}})d\xi$.
Both $I$ and $\sigma_{I_{i}}^{2}=\sum_{i}V_{ij}^{2}\sigma_{nz}^{2}$
depend on $V_{ij}$, and: 
\begin{align}
\partial_{V_{ij}}M & =\sum p_{x}\{x_{j}\partial_{I}G+\partial_{V_{ij}}\sigma_{I}\partial_{\sigma}G\}\bigg\{\ln\frac{G(Vx,|V\sigma_{nz}|^{2})}{1-G(Vx,|V\sigma_{nz}|^{2})}-\ln\frac{p(y_{i}=1)}{1-p(y_{i}=1)}\bigg\}\nonumber \\
 & =\sum p_{x}\{x_{j}\partial_{I}G+\frac{V_{ij}\sigma_{j}^{2}}{\sigma_{I}}\partial_{\sigma_{I}}G\}\bigg\{\ln\frac{G(Vx,|V\sigma_{nz}|^{2})}{1-G(Vx,|V\sigma_{nz}|^{2})}-\ln\frac{p(y_{i}=1)}{1-p(y_{i}=1)}\bigg\}
\end{align}
now contains an additional term proportional to $\frac{V_{ij}\sigma_{j}^{2}}{\sigma_{I}}$
when compared to the noiseless case. This term is usually negative,
meaning that larger synaptic weights increase uncertainty. It also
does not vanish for $x_{j}=0$, effectively acting as a leak term
for the synaptic weight dynamics.

For the particular choice of $f(I)=H(I-\theta)$ as a step function,
one gets $G(I,\sigma)=1-\Phi(\frac{\theta-I}{\sigma})=0.5-0.5erf(\frac{\theta-I}{\sqrt{2}\sigma})$
and $\partial_{\sigma}G=\frac{\theta-I}{\sqrt{2\pi}\sigma^{2}}\exp(\frac{-(\theta-I)^{2}}{2\sigma^{2}})=\frac{\theta-I}{\sigma}\partial_{I}G$.
(Additionally, some stochasticity in the neuronal output can be modelled
by adding the constant $\sigma_{y}^{2}$ to $\sigma_{I}^{2}$.)

To obtain a learning rule operating with $\tilde{x}=x+\xi$ instead
of $x$, if the network only receives noisy itput, one can rewrite
$\partial_{V}M$ as: 
\begin{align}
\partial_{V}M & =\sum p(\tilde{x})\tilde{x}f^{/}(V\tilde{x})\bigg\{\int p(\xi)\ln\frac{G(\tilde{x}-\xi,\sigma)}{1-G(\tilde{x}-\xi,\sigma)}d\xi-\ln\frac{p(y=1)}{p(y=0)}\bigg\}\nonumber \\
 & +\sum p(\tilde{x})p(y|\tilde{x})\bigg\{\int p(\xi)\bigg[(\tilde{x_{j}}-\xi_{j})\partial_{I}+\frac{V_{ij}\sigma_{j}^{2}}{\sigma_{I}}\partial_{\sigma_{I}}\bigg]\ln p(y|(V(\tilde{x}-\xi),\sigma_{I}^{2}))d\xi\bigg\}
\end{align}
with $p(y|\tilde{x})=[1-y+f(V\tilde{x})(2y-1)]$. For a particular
form of $f$ and $G$, integration over $\xi$ can be performed to
get a new learning rule. For the above-mentioned step function, one
can exploit the relation $\partial_{\sigma}G=\frac{\theta-I}{\sigma}\partial_{I}G$.
Considering input from channel $j$ and from all other channels in
neuron $i$ separately, one obtains by integration in the leading
order of $\sigma_{nz}$ an approximation for $\partial_{V_{j}}M\sim p(\tilde{x})\{p^{/}(y|\tilde{x})\tilde{x}-V_{j}\sigma_{j}^{2}/\sigma_{I}^{2}p(y|\tilde{x})\}\ln\frac{p(y|\tilde{x})}{p(y)}$
with a new noise-induced term proportional to $-V_{j}\sigma_{j}^{2}/\sigma_{I}^{2}$.

\section*{Acknowledgements}

This work was supported by the Swiss National Science Foundation (31003A\_170079)
and by the Australian Research Council (DP180101494) to RBJ.

\end{document}